\documentclass[
prl,
twocolumn,
amsmath,
amssymb,
aps,
]{revtex4-2}

\usepackage{graphicx}
\usepackage{dcolumn}
\usepackage{bm}
\usepackage{braket}
\usepackage[dvipsnames]{xcolor}
\usepackage{color}

\newcommand{\blue}{\textcolor{black}}

\usepackage{hyperref}
\usepackage{soul} 
\def\3he{$^3$He} 
\def\4he{$^4$He}

\def\Aph{\textit{A} phase}
\def\Bph{\textit{B} phase}

\def\pwave{\textit{p}-wave}

\definecolor{tabgreen}{RGB}{44, 160, 44}
\definecolor{tabpurple}{RGB}{148, 103, 189}
\definecolor{taborange}{RGB}{255, 127, 14}

\begin{document}

\preprint{}

\title{Phase Stability of Superfluid \3he in Anisotropic Aerogel}

\author{J. W. Scott}
\email{johnscott2025@u.northwestern.edu}
\author{D. Park}
\author{X. Yuan}
\author{W. P. Halperin}
\email{w-halperin@northwestern.edu}
\affiliation{Northwestern University, Evanston, IL 60208, USA}

\date{\today}

\begin{abstract}
	The \textit{A} and \textit{B} phases of superfluid \3he{}  have vector degrees of freedom that reflect their characteristic broken symmetries, respectively chiral and spin-orbit rotation axes.
	Anisotropic disorder in the superfluid, imbibed in uniformly strained silica aerogel, orients these degrees of freedom, thereby affecting phase stability.
	These degrees of freedom have been found to spontaneously reorient at a field-independent transition temperature $T_{x}$, that can be accounted for with a temperature dependent anisotropic Ginzburg-Landau model.

\end{abstract}

\maketitle

In its natural state, liquid \3he{} is well-established  as a spin-triplet \pwave{} superfluid. As such it is ideal for  investigation of the effect of surfaces and impurities, relevant to understanding putative odd-parity superconducting materials.
The \pwave\,\,triplet order parameter possesses vector orientational degrees of freedom, the orbital chiral axis $\bm{\ell}$ in the \Aph{} and spin-orbit rotation axis $\bm{\hat{n}}$ in the \Bph{}, similar to putative odd-parity superconductors\cite{Vol.13,Aok.22,Kin.23}.
Introducing anisotropic impurity in the form of uniformly-strained aerogel systematically reduces the symmetry of the orbital pairing potential, splitting it into components aligned with and perpendicular to the strain axis $\bm{\hat{\epsilon}}$\cite{Aoy.06, Sau.13}.
The orbital pairing potential couples directly to the spin degrees of freedom through the superfluid dipole-dipole interaction  producing a nuclear magnetic resonance (NMR) frequency shift\cite{Leg.73}. Here we  develop an anisotropic Ginzburg-Landau (GL) model to interpret measured frequency shifts that identify the orientation of the vector degrees of freedom  attributable to  aerogel anisotropy.

\begin{figure}
	\includegraphics[width=\columnwidth]{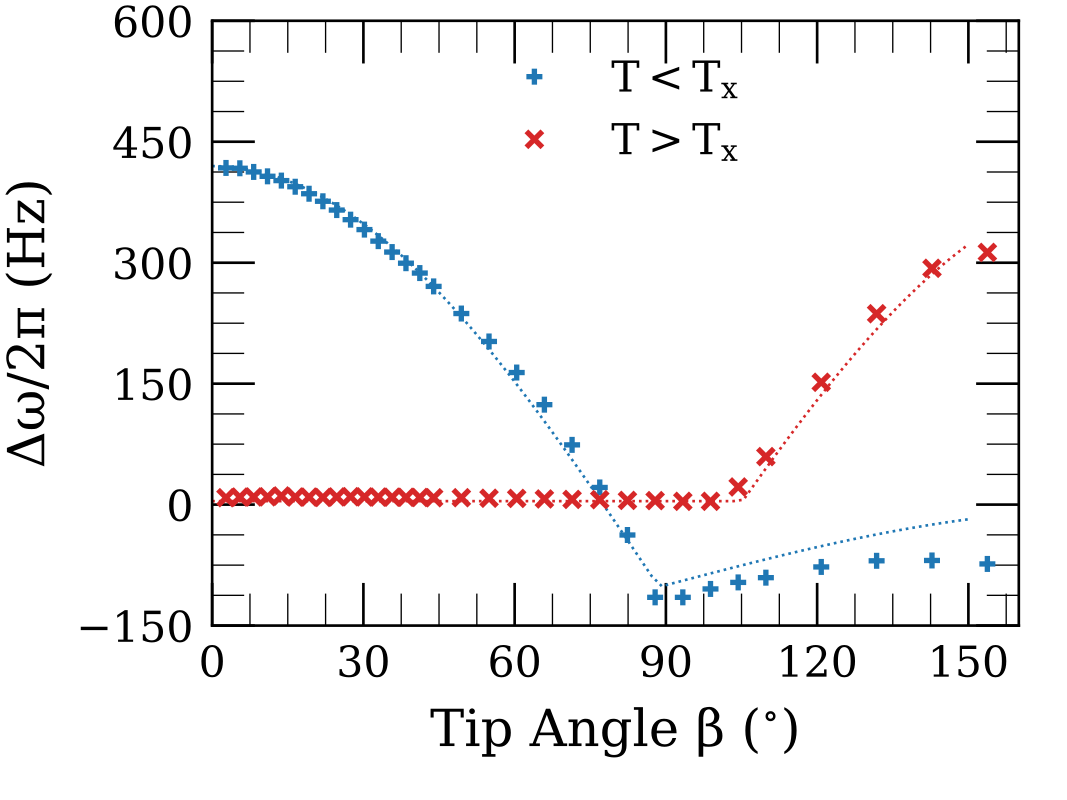}
	\caption{\label{fig:tip_angle_shifts}
	Tip angle dependence of NMR precession frequency shift.  The two possible modes are $\bm{\hat{n}}\parallel\bm{H_{0}}$ (red) and $\bm{\hat{n}} \nparallel \bm{H_{0}}$ (blue) above and below $T_{x}$ at $T/T_{c} = 0.71$ and $T/T_{c} = 0.84$ temperatures respectively, in $16\%$ positive strain aerogel\cite{Zim.19}. Crosses are experimental data at $P = 26.5$ bar, $H_{0} = 74.4$ mT. Dashed lines represent theoretical calculations incorporating order parameter distortion\cite{Has.83,Dmi.14}.
}
\end{figure}

\begin{figure}
\includegraphics[width=\columnwidth]{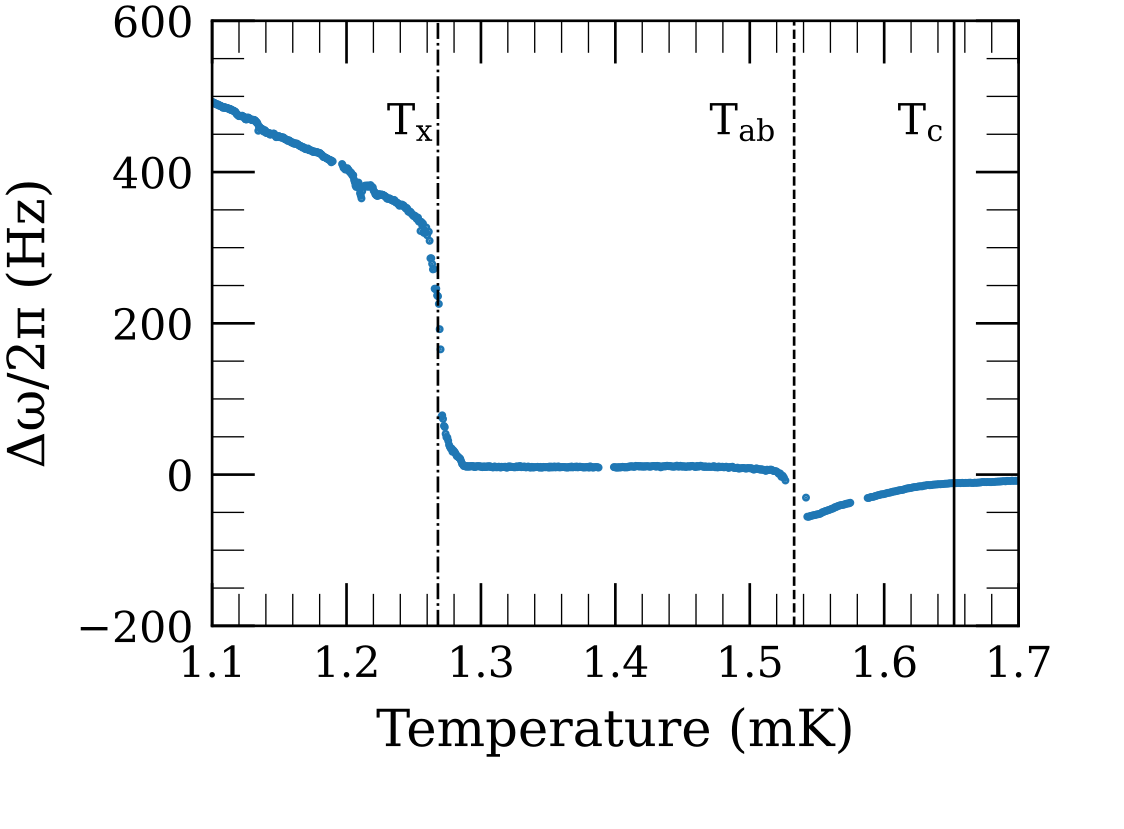}
	\caption{\label{fig:warmup_freq_shift}Temperature dependence of the NMR resonance frequency shift in $16\%$ positive strain aerogel in the \Bph{} at $P = 26.5$ bar, $H_{0} = 74.5$ mT at tip angle $\beta \approx 0$.
	The transition $T_{x}$ from $\bm{\hat{n}} \nparallel \bm{H_{0}}$ to $\bm{\hat{n}} \parallel \bm{H_{0}}$ occurs near $1.26$ $\mathrm{mK}$.
	}
\end{figure}

We have  reported that  the measured NMR frequency shifts of the superfluid  in anisotropic aerogel exhibit a sharp reorientation transition as a function of temperature at $T_{x}$ as shown in Fig.\ref{fig:tip_angle_shifts} and Fig.\ref{fig:warmup_freq_shift}, previously described as an orbital flop transition in the \Aph{}, and more appropriately in the \Bph{} as a $\bm{\hat{n}}$ flop transition\cite{Zim.18,Sco.23,Ngu.24}.
These experiments, complemented by our new work with non-magnetic surface scattering and positive strained aerogel, have motivated a phenomenological model for this transition, reported here.
This extends our understanding of phase and textural stability for the experimental parameter space that includes the region of unexpected enhancement of the magnetic susceptibility in the \Bph{}\cite{Sco.23}.
They  are also relevant to the reorientation transition of the chiral axis in the \Aph{}, important for detection of the predicted anomalous thermal Hall effect in impure chiral superfluids\cite{Nga.20,Sha.22}.
In the \blue{end matter section} we discuss the effects of solid \3he{} that forms on the surface\cite{Spr.96,Col.09}.

The transition in the orientation of these vector degrees of freedom is identified through the nuclear magnetic resonance (NMR) spectrum of superfluid \3he{}.
Macroscopic coherence in superfluid \3he{} amplifies the nuclear dipole-dipole interaction and induces a  NMR frequency shift $\Delta \omega(\beta)$. 
The frequency shift dependence  on tip angle $\beta$, via the NMR pulse length, shown in Fig.\,\ref{fig:tip_angle_shifts} provides a precise measure of the structure of the superfluid order parameter $d_{\mu j}$\,\cite{Leg.73,Leg.74,Bri.75a,Bri.75b}.
In the pure \Bph{}, spontaneously broken spin-orbit symmetry is expressed as  a relative rotation of the spin and orbit vectors about the $\bm{\hat{n}}$ axis, resulting in an order parameter of the form $d_{\mu j} = e^{i\phi} (\Delta_{B} / \sqrt{3}) R_{\mu \nu}(\bm{\hat{n}}, \theta) \delta_{\nu j}$, with $\theta \approx \cos^{-1}(-1/4)$.
$R(\bm{\hat{n}},\theta)$ is the rotation operator and ${\mu,  j}$ spin and orbit coordinate indices.
In strained aerogel, there are two orientations of the vector $\bm{\hat{n}}$ with respect to the static magnetic field that correspond to two different modes of \3he{} spin precession, evident  in Fig.\,\ref{fig:tip_angle_shifts}.
Depending on temperature, either orientation is observed  with a transition between them at temperature $T_{x}$. 

In the \Bph, $\bm{\hat{n}}$ is oriented by a combination of surface and field effects. 
For magnetic fields larger than the dipole field, $H_{D} \approx 3$\,mT,  and at distances far from a surface, the effect of order parameter distortion on the dipole energy weakly favors the orientation $\bm{\hat{n}} \parallel \bm{H_{0}}$;  $\bm{H_{0}}$ being the static magnetic field of the NMR experiment.
Near a surface with surface normal $\bm{\hat{s}}$, aligned such that  $\bm{\hat{s}}\bot\bm{H_{0}}$, pair breaking suppresses the orbitals perpendicular to the surface leading to a Zeeman energy minimized when $\bm{\hat{n}}$ is oriented at an angle $~63^{\circ}$ away from the magnetic field\cite{Bri.74b,Osh.75}. 
Similarly, in the pure \Aph{} a surface $\bm{\hat{s}}$ orients  $\bm{\ell}\parallel\bm{\hat{s}}$, with $\bm{\ell}$ being the chiral axis and nodal direction\cite{Amb.74}.
The anisotropic susceptibility of the \Aph{} together with the dipole energy weakly reorients $\bm{\ell}\bot\bm{H_{0}}$ in order to minimize the dipole energy.

\begin{figure}
	\includegraphics[width=\columnwidth]{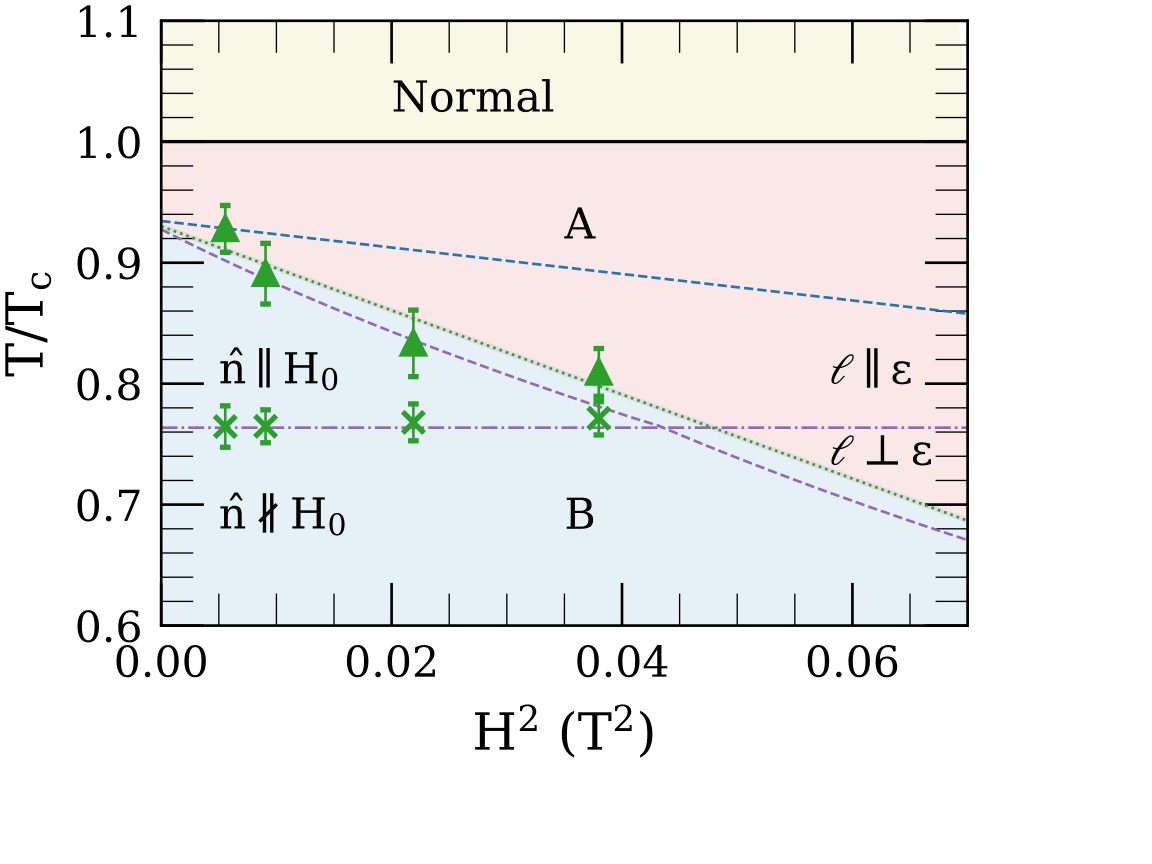}
	\caption{\label{fig:phase_diagrams}
	Phase Diagrams for $T_{AB}$ and flop temperature $T_{x}$ dependence on magnetic field in $16\%$ positive strain aerogel at 26.5 bar, experiment and theory, with the A-phase shaded red and B-phase blue.
	A fit to the $H^{2}$ dependence of experimentally measured $T_{AB}$ is given by the green dotted line.
	$T_{AB}$ with weak coupling $\beta_{i}$ and $T_{c,\parallel} / T_{c,\bot} = 0.96$ is shown with the blue line\cite{Thu.98}. 
	Strong-coupling $\beta_{i}$ with $T_{c,\parallel} / T_{c,\bot} = 0.99$ is given by the purple dashed line\cite{Thu.98,Wim.18,Reg.20}.
	The flop transition temperature $T_{x}$ is shown with a purple dash-dot line.
	Experimental $T_{AB}$ is shown with the green triangles, $T_{x}$ with green crosses.
	}
\end{figure}

These changes in orientation of the $\bm{\hat{n}}$ and $\bm{\ell}$  vectors, and concomitant changes in dipole energy of the superfluid, produce different NMR frequency shifts. 
In Fig.\ref{fig:tip_angle_shifts} the frequency shifts of two different orientations of the $\bm{\hat{n}}$-vector of the \Bph{}, identified from their $\Delta \omega$ dependence on $\beta$, at temperatures above and below $T_{x}$ are shown, a consequence of reorientation of $\bm{\hat{n}}$ \cite{Zim.19}.
We have observed the $\bm{\hat{n}}$ reorientation transition $T_{x}$ in the \Bph{} in both positive and negative strained aerogels, also previously referred to as stretched and compressed respectively. 
The $T_{x}$ transition is sharply defined and uniform throughout the sample as shown in Fig.\,\ref{fig:warmup_freq_shift}, although NMR measurements of the transition are spatially averaged over a dipole length $\xi_{D} \approx 7-10$ $\mu$m\cite{Zim.18,Sco.23}.
The transition temperature was found to be magnetic field-independent over a range of fields, as shown in Fig.\,\ref{fig:phase_diagrams} for positive strain.
In negatively strained aerogels, the orientations of the $\bm{\hat{n}}$\,vector above and below $T_{x}$ are opposite to that for positive strain \cite{Zim.18}.

Strained aerogel introduces anisotropic quasiparticle scattering distorting the order parameter of the \Bph{} to 
\begin{equation}
	\label{eq:dist_bph}
	d_{\mu j} = e^{i\phi}\frac{R_{\mu \nu}(\bm{\hat{n}},\theta)}{\sqrt{3}}(\Delta_{\bot}(\hat{x}_{\nu}\hat{x}_{j} + \hat{y}_{\nu}\hat{y}_{j}) + \Delta_{\parallel} \hat{z}_{\nu}\hat{z}_{j}),
\end{equation}
with $\Delta_{\parallel}$ and $\Delta_{\bot}$ being the order parameter amplitudes, respectively parallel and perpendicular to the strain axis. 
In the present work, with positive strain and $\bm{\hat{\epsilon}}\parallel \bm{H_{0}}$, there are  two possible equilibrium orientations of $\bm{\hat{n}}$, separated in temperature at $T_{x}$. 
On cooling from a planar-distorted \Bph{} with distortion $\Delta_{\parallel} / \Delta_{\bot} < 1$, the transition is to a polar-distorted \Bph{} with $\Delta_{\parallel} / \Delta_{\bot} > 1$.
In each case, the distortion of the order parameter amplitudes and the orientation of the $\bm{\hat{n}}$ vector determine the NMR frequency shift.

When $H_{0} > H_{D}$, $\bm{\hat{n}}$ is oriented such that $d_{z j}$ has an amplitude determined solely by the lesser of $\Delta_{\bot}$ or $\Delta_{\parallel}$.
This ensures that the $M_{z} = 0$ pairs, with no magnetization parallel to $H_{0}$, exist in the more suppressed orbital, minimizing the Zeeman energy.
When $\Delta_{\parallel} < \Delta_{\bot}$, this effect in conjunction with the dipole energy $\bm{\hat{n}}\parallel\bm{H_{0}}$, similar to the bulk fluid.
Whereas, when $\Delta_{\parallel} > \Delta_{\bot}$ this orients $\bm{\hat{n}} \nparallel \bm{H_{0}}$ similar to the wall-pinned mode.
Fitting the $\beta$ dependence of the frequency shifts, shown in Fig.\ref{fig:tip_angle_shifts}, determines the order parameter distortion $\Delta_{\parallel} / \Delta_{\bot}$\cite{Has.83,Dmi.14,Ran.96}. 
For the $T > T_{x}$ data in Fig.\ref{fig:tip_angle_shifts}, a fit to the $\bm{\hat{n}} \parallel \bm{H_{0}}$ mode gives $\Delta_{\parallel}/\Delta_{\bot} = 0.94\pm0.01$,
In the $T < T_{x}$ data, a fit to the $\bm{\hat{n}} \nparallel \bm{H_{0}}$ mode gives $\Delta_{\parallel} / \Delta_{\bot} = 1.08 \pm 0.03$.
The suppression of one of the $M_{z} = 0$ orbitals in the polar-distorted case, while not shown in Eq.\ref{eq:dist_bph} is important to the thermodynamics of the flop transition.
The full form of the polar-distorted order parameter, as well as the \blue{effect} of order parameter distortion on the NMR frequency shift are considered in \blue{the end matter section}.

In order to describe  the transition between these distorted states, we adopt a phenomenologically-motivated GL approach.
For pure superfluid \3he{} this is exceptionally robust, based on material parameters and theoretical models for microscopic interactions in good agreement with experiment\cite{Mer.73,Cho.07,Wim.18}.
Similarly, an adaption to the impure superfluid has been  successful in predicting the stabilization of the isotropic state by isotropic quasiparticle scattering\cite{Thu.98,Ger.02}.
Quenched disorder from the random structure of aerogel also disorders the vector degrees of freedom of the superfluid, observed in both isotropic and anisotropic aerogels\cite{Li.13,Li.14b,Dmi.10}.
Straining the aerogel, produces structural anisotropy  that competes with  disorder establishing long-range order of the vector degrees of freedom\cite{Aoy.06,Vol.08,Sau.13,Sur.19}.
The structural anisotropy splits the orbital pairing amplitudes of the anisotropic GL free energy functional,
\begin{multline}
	\label{eq:f_GL}
	F - F_{N} = \alpha_{\bot}(d^{*}_{\mu x} d_{\mu x} + d^{*}_{\mu y} d_{\mu y}) \\ + \alpha_{\parallel}(d^{*}_{\mu z} d_{\mu z})  + \beta_{1} |d_{\mu i} d_{\mu i}|^{2} + \beta_{2} (d_{\mu i}^{*} d_{\mu i})^{2} \\ + \beta_{3}(d_{\mu i}^{*} d_{\nu i}^{*} d_{\mu j} d_{\nu j}) + \beta_{4}(d_{\mu i}^{*} d_{\nu i} d_{\nu j}^{*} d_{\mu j})\\ + \beta_{5}(d_{\mu i}^{*} d_{\nu i} d_{\nu j} d_{\mu j}^{*}) + g_{Z}(H_{\mu} d^{*}_{\mu i} d_{\nu i} H_{\nu}),
\end{multline}
where $\alpha_{\bot,\parallel}$ are the quadratic-order GL invariants for each orbital pairing potential while the parameters $\beta_{i}$ are the fourth-order GL invariants, $g_{Z}$ is the Zeeman energy coefficient.
$H_{\mu}$ is taken to be the static NMR field $H_{0}$.
In the impure superfluid all \blue{of} these material parameters depend on the pairbreaking parameter $x = \hbar v_{F} / (2 \pi \lambda k_{B} T)$ which captures the effects of impurity scattering on the superfluid with the influence of impurity scattering entering via the comparison of the Fermi velocity $v_{F}$ and $\lambda$, the transport mean-free path\cite{Sau.03}.

Substituting the order parameter for the planar and polar-distorted \Bph{} decribed by Eq.\ref{eq:dist_bph} into the equation for the GL free energy in Eq.\ref{eq:f_GL} we find that they are equivalent if $\alpha_{\bot} = \alpha_{\parallel}$.
Notably, this equivalence is independent of both the values of the $\beta_{i}$ parameters as well as the magnetic field.
The temperature $T_{x}$ is therefore defined by a point where $\alpha_{\bot} = \alpha_{\parallel}$, while above and below, the orientation of $\bm{\hat{n}}$ vector is defined by which of the two quadratic-order terms is greater.

The $\bm{\hat{n}}$-flop transition has been observed in both positive and negative strained aerogels. Here we only address the positive strain case where above $T_{x}$ $|\alpha_{\bot}| > |\alpha_{\parallel}|$\cite{Sco.23}.
Extending the model to negative strain is straightforward, requiring the exchange of the relative amplitudes of $\alpha_{\parallel, \bot}$ near $T_{x}$, consistent with the identification of anisotropic aerogel as falling into two classes defined by which of the pairing channels  $\alpha_{\parallel}$ or $\alpha_{\bot}$ are favored close to $T_{c}$. 
For negative strain this leads to nematic-like pairing discussed by Nguyen $et\,\,al.$\cite{Ngu.24}.

We make the ansatz that the experimentally measured $T_{c}$ is $T_{c,\bot}$ for the $\alpha_{\bot}$ pairing channel taking the amplitudes of $\alpha_{\bot}(T)$ and $\alpha_{\parallel}(T)$ to be linear in temperature, \textit{i.e.} $\alpha_{\bot,\parallel}(T) = \alpha_{\bot,\parallel}'(T/T_{c\,\bot,\parallel} - 1)$.
This phenomenological approach does not require a specific prediction for the form of the pairbreaking parameter for the pairing channel governed by $\alpha_{\parallel}$.
This leaves two free parameters determining $\alpha_{\bot,\parallel}(T)$, $T_{c\parallel}$ and $\alpha(T_{x}) = \alpha_{\bot}(T_{x}) = \alpha_{\parallel}(T_{x})$.
The fourth-order $\beta_{i}$ have their weak-coupling values adjusted per the theory for the impure superfluid, assuming a random phase shift $\sin^{2}{\delta_{0}} = \frac{1}{2}$\cite{Thu.98}.
Strong coupling corrections of the form $\beta_{i} = \beta_{i}^{wc}(1 + \delta \beta_{i}^{sc}(T/T_{c0}))$, with $T_{c0}$ the pure superfluid $T_{c}$, were used with impurity-adjustment to the weak-coupling $\beta_{i}$ given by $\delta \beta_{i}^{sc}$ for the pure superfluid\cite{Wim.18,Reg.20}.
Similarly, the Zeeman coupling constant $g_{Z}$ is corrected to its impure value\cite{Sha.01}. 

The measured NMR signal is limited in spatial resolution by the dipole length $\xi_{D} \approx 7-10 \,\mu\textrm{m}$. 
In constructing the relationship between the measured NMR signal and the GL theory, we take the order parameter to be well-described by a homogeneous average order parameter $\langle d_{\mu j} \rangle_{\xi_{D}}$.
Spatial variations in the order parameter, such as seen in the Larkin-Imry-Ma effect, are neglected in this description\cite{Vol.08,Sau.13,Li.13}.
Contributions to the GL free energy linear in the magnetic field, responsible for the splitting of $T_{c}$ between different equal-spin-pairing channels, are also neglected\cite{Ger.02,Sau.03,Sur.19,Dmi.21}.

\begin{figure}
\includegraphics[width=\columnwidth]{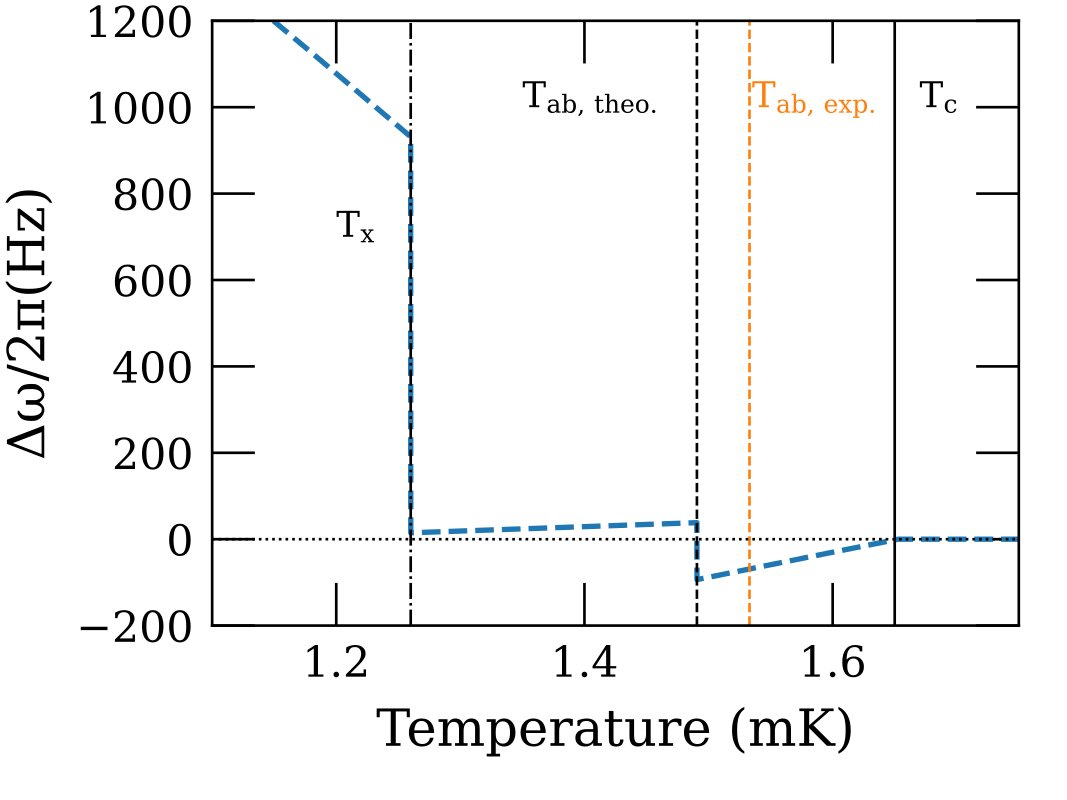}
	\caption{\label{fig:theory_freq_shifts}
	Calculated small tip angle ($\beta = 0$) NMR frequency shifts $\Delta \omega / 2 \pi$ as a function of temperature for the phase diagram at $H_{0} = 0.0745 \, \textrm{T}$, $P=26.6$ bar, strong coupling $\beta_{i}$.
	The experimentally measured $T_{ab}$ is shown in orange, while the calculated $T_{ab}$ is shown in black.
	The linear extrapolation of the GL coefficent $\alpha$ to $\alpha(T_{x})$ based on the pressure dependence of $T_{c}$ over estimates the frequency shift at low temperatures.
	}
\end{figure}

The model for the pairbreaking parameter $x$ assumes a homogeneous and isotropic distribution of impurities.
Correlations in the aerogel impurity network give additional corrections to the suppression of the order parameter amplitude in the superfluid\cite{Thu.98,Fom.08}.
A simple correction to the pairbreaking parameter, proposed by Sauls and Sharma, adequately describes the pressure dependence of $T_{c}$ in most silica aerogel samples published to date\cite{Sau.03}.
In this case, the pairbreaking parameter takes a form $x_{IISM} = (x)/(1 + \xi_{a}^{2}/( \lambda^{2} x))$, where $\xi_{a}$ is a correlation length in a model that interpolates between  homogeneous-scattering at low pressures, where the pure superfluid coherence length $\xi_{0}$ is comparable to $\lambda$, and at high pressures  where $\xi_{0}$ is smaller and consequently  $\xi_{a}$ is determined by random voids in the aerogel.
The value for $\alpha(T_{x})$ is extrapolated from $\alpha'(T_{c})=\alpha_{\bot}^{'}$ using the $x_{IISM}$ from the pressure dependence of $T_{c}$\cite{Sco.23}.
With these assumptions, we can calculate both a magnetic field phase diagram, choosing $T_{c,\parallel}$ such that the calculated $T_{AB}$ at $H_{0} = 0$ is close its extrapolated experimental value, as shown in Fig.\ref{fig:phase_diagrams}.
Similarly, by assuming the dipole coupling constant $g_{D}$, which determines the scale of the NMR frequency shift is the same as its pure value, we can then calculate a temperature dependent NMR frequency shift, shown in Fig.\ref{fig:theory_freq_shifts}\cite{Thu.00}.

\begin{figure}
	\includegraphics[width=\columnwidth]{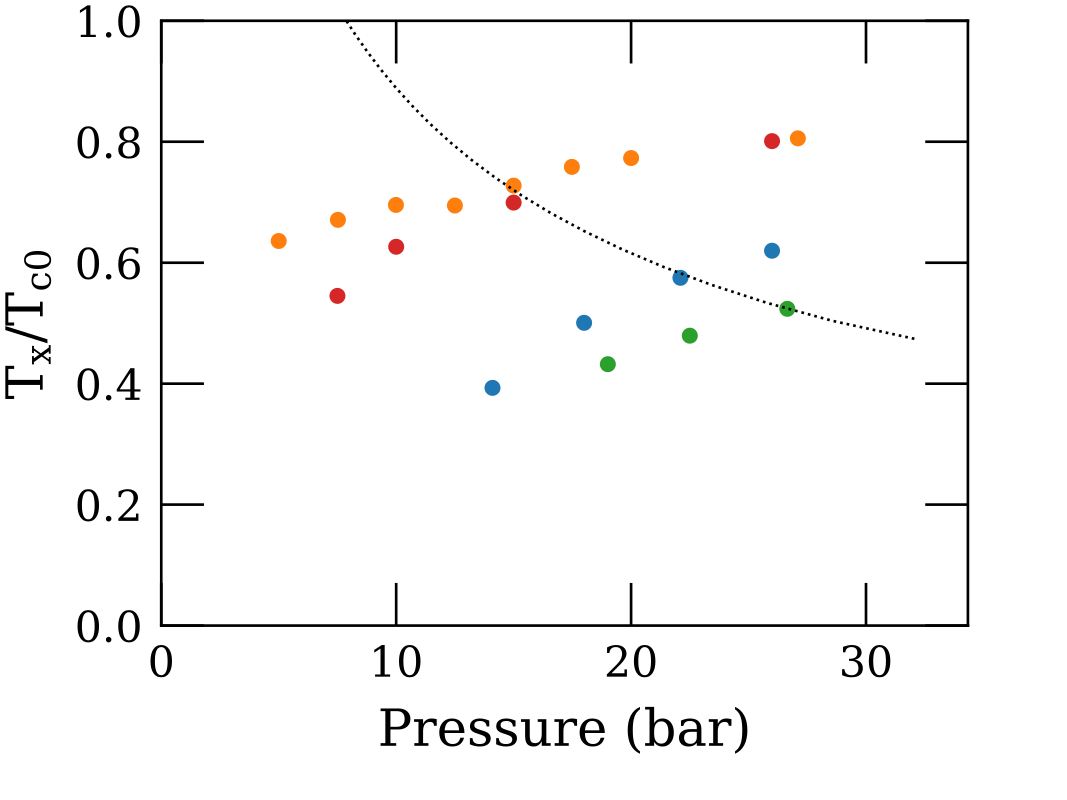}
	\caption{\label{fig:pressure_tx_dependence}
	Pressure $P$ dependence of $T_{x}$ in preplated and non-preplated positive strain (green preplated $16\%$ positive strained, blue non-preplated $14\%$ positive strain) and negative strain (red non-preplated, orange preplated) aerogels\cite{Pol.12,Zim.18,Zim.19t}.
	The pure coherence length, appearing in Eq.\ref{eq:tx_iism_pressure} is a well-established function of pressure\cite{Hal.90}.
	The positive slopes $d(T_{x}/T_{c0}) / d P$ are not compatible with the pair breaking parameter $x$ for either the homogeneous or inhomogeneous scattering models\cite{Thu.98,Sau.03}.
	The black dotted line corresponds to the prediction for the anisotropic IISM model given by equation Eq. \ref{eq:tx_iism_pressure} for the data point taken in preplated $16\%$ positively strained aerogel at $P \approx 26.5,\mathrm{bar}$.
}
\end{figure}

The condition for the flop $\alpha_{\bot}(T_{x}) = \alpha_{\parallel}(T_{x})$ necessitates that at $T_{x}$ $x_{\bot} = x_{\parallel}$.
If we were to assume both quadratic-order terms $\alpha_{\bot}$ and $\alpha_{\parallel}$ are described by $x_{IISM}$ with different $\lambda_{\bot}, \lambda_{\parallel}$ and $\xi_{a,\bot}, \xi_{a,\parallel}$, a solution to $x_{\bot} = x_{\parallel}$ exists at temperature
\begin{equation}
	\label{eq:tx_iism_pressure}
	\frac{T_{x}}{T_{c0}} = \xi_{0}(P) \frac{\lambda_{\bot} - \lambda_{\parallel}}{\xi_{a,\parallel}^{2} - \xi_{a,\bot}^{2}}.
\end{equation}
This predicts, based on the pressure dependence of the pure coherence length $\xi_{0}(P)$, that $T_{x}/T_{c0}$ should increase at low pressure.
The opposite pressure dependence is observed experimentally, as shown in Fig.\ref{fig:pressure_tx_dependence}, implying the need for higher-order corrections to the functional form of the pairbreaking parameter for anisotropic silica aerogels.
Evidence for the existence of a crossover in the aerogel transport free paths and particle-particle correlation length appears in biased diffusion limited cluster aggregation (DLCA) models,\cite{Ngu.24} and may provide the basis for an additional correction to the inhomogeneous anisotropic pair breaking parameter.
Notably in highly anisotropic aerogels, no $T_{x}$ is observed suggesting that the reduced flop transition temperature $T_{x}/T_{c0}$ might be completely suppressed\cite{Dmi.20a,Dmi.20b}.

An additional consequence of this crossover in pairing is a flop transition of the chiral axis $\bm{\ell}$ in the \Aph{} at the same $T_{x}$ as in the \Bph{}, as shown in Fig.\ref{fig:phase_diagrams}.
Extrapolating the $AB$ transition to higher magnetic fields suggests that this transition should occur in positive strain aerogels at fields above roughly $0.22\,\textrm{T}$.
A flop in the \Aph{} similar to that described here has been  observed in positive strain aerogel without \4he{} preplating\cite{Pol.12,Li.14c}.
The reason for the favorability of the \Aph{}, which is the only superfluid phase observed to the lowest temperatures achieved in these experiments, in the presence of solid paramagnetic \3he{} covering the aerogel surface is not known.

Study of the temperature-dependent crossover in the orbital pairing potential can be performed with existing anisotropic Ginzburg-Landau theories of superfluid \3he{} imbibed in anisotropic aerogel reproducing order-parameter orientational transitions measured with nuclear magnetic resonance experiments.
In future experimental work the reorientation of the chiral axis of the \Aph{} at high-fields will be investigated important for an unambiguous detection of the impurity-mediated anomalous thermal hall effect in \3he{} in aerogel\cite{Sha.22}.
NMR measurements of $\mathrm{UTe}_{2}$ subject to hydrostatic pressure show a sharp change in the Knight shift at a second transition below $T_{c}$, indicative of rotation of the spin degrees of freedom of a candidate \pwave{} superconductor\cite{Kin.23,Hay.25,Wan.25}.

While our model is successful in describing the observed flop transition in superfluid \3he{} imbibed in \4he{} preplated aerogel, it is not compatible with experimentally observed phenomena in non-preplated aerogel, discussed in more detail in \blue{the end matter section}.

This work was supported by NSF Division of Materials Research Grant No. DMR-2210112. We are grateful to M. D. Nguyen, J. A. Sauls, and A.B. Vorontsov for useful discussion on this work.

\bibliography{Scott_References}

\section{End Matter}

\subsection{Magnetic Surfaces}
In contrast to measurements performed in nanofluidic confinement, $T_{c}$ suppression is approximately the same for \3he{} imbibed in \4he{} preplated versus non-preplated aerogel\cite{Zim.18,Hei.21}.
However, despite the small change in $T_{c}$, preplating exerts a strong influence on phase stability.
For positively strained silica aerogels without \4he{} preplating  the \Bph{} is supressed for all measured fields and temperatures\cite{Pol.12}.
In this case a separate flop transition occurs where the $\bm{\ell}$-vector spontaneously reorients from parallel to perpendicular to the strain axis $\hat{\epsilon}$ at a comparable temperature $T_{x}$\cite{Li.14c,Ngu.24}.
This measurement prompted the calculation of a biaxial phase in superfluid \3he{} exhibiting a similar continuous reorientation of the $\ell$ vector, which could potentially exist between $T_{c\parallel}$ and $T_{x}$ in this model\cite{Sau.13}.
The existence of such a flop transition in strained aerogel, together with the stability of the \Bph{} in preplated stretched aerogel, implies that the mechanism for the anomalous observed stability of the \Aph{} in unpreplated stretched aerogel is not suppression of an orbital pairing amplitude but rather a higher-order effect related to the presence of magnetic impurity.
It has further been measured that in negatively-strained aerogel the \Bph{} remains stable to a nonzero critical field $H_{c}$\cite{Li.14b}.
Some adjustment to the effective anisotropy of the pairing potential is apparent in the adjustment of the flop temperature $T_{x}$ seen in Fig.\ref{fig:pressure_tx_dependence}, possibly consistent with a microscopic approach to anisotropic exchange scattering\cite{Min.18}.
Understanding the influence of exchange scattering, however, may require extension to higher-order GL terms similar to treatments in the isotropic case\cite{Sau.03,Bar.04}

\subsection{Distortion and NMR Frequency Shifts}

The planar distorted order parameter described by Eq.\ref{eq:dist_bph} in the case $\Delta_{\bot} > \Delta_{\parallel}$ has a frequency shift of form
\begin{widetext}
\begin{equation}
	\label{eq:planar_dist_freq_shift}
	\omega_{L} \Delta \omega(\beta) = \\ \begin{cases}
		\frac{g_{D}}{3} \frac{\gamma^{2}}{\chi} (\Delta_{\bot}^{2} - \Delta_{\parallel}^{2}) &  \beta < \beta^{*}\\
		\frac{g_{D}}{3} \frac{\gamma^{2}}{\chi}(\Delta_{\bot} + \Delta_{\parallel})(\Delta_{\bot} + 2 (\Delta_{\bot} + \Delta_{\parallel})\cos{\beta}) & \beta > \beta^{*} \\
		\end{cases}
\end{equation}
\end{widetext}
	for critical tip angle $\beta^{*}$ that satisfies
\begin{equation}
	\cos{\beta^{*}} = \frac{-2 \Delta_{\bot} + \Delta_{\parallel}}{2(\Delta_{\bot} + \Delta_{\parallel})}.
\end{equation}
determining the tip angle which separates the two modes of precession\cite{Has.83}.
When polar-distorted, the order given in Eq. \ref{eq:dist_bph} does not include the influence of the Zeeman energy on the order parameter.
The suppression of $M_{z} = 0$ pairs further suppresses an order parameter amplitude in the planar direction.
Without loss of generality, we can choose the Zeeman-suppressed component to be in the $\hat{x}_{j}$ orbital direction, leading to an order parameter
\begin{equation}
	\label{eq:polar_dist_bph}
	d_{\mu j} = e^{i\phi}\frac{R_{\mu \nu}(\bm{\hat{n}},\theta)}{\sqrt{3}}(\Delta_{Z, \bot} \hat{x}_{\nu}\hat{x}_{j} + \Delta_{\bot} \hat{y}_{\nu}\hat{y}_{j} + \Delta_{\parallel} \hat{z}_{\nu} \hat{z}_{j})
\end{equation}
where is $\Delta_{Z,\bot}$ is the Zeeman-suppressed amplitude of the planar component of the order parameter.
When the static magnetic field is zero, $\Delta_{Z,\bot} = \Delta_{\bot}$.
For this order parameter, a variety of modes of precession may be stabilized for highly distorted states, but the modes observed experimentally are described by
\begin{widetext}
\begin{equation}
	\label{eq:planar_dist_freq_shift}
	\omega_{L} \Delta \omega(\beta) = \\ \begin{cases}
		\frac{g_{D}}{3} \frac{\gamma^{2}}{\chi} (\frac{\Delta_{\bot}^{2}}{4}(1-\cos{\beta}) + (\frac{\Delta_{\bot,Z}^{2}\Delta_{\parallel}^{2}}{\Delta_{\bot}^{2}}(2\cos{\beta} - 1)+(\Delta_{\bot,Z}^{2} + \Delta_{\parallel}^{2})\cos{\beta})) &  \beta < \beta^{*}\\
		\frac{g_{D}}{3} \frac{\gamma^{2}}{\chi} (\frac{-\Delta_{\bot}^{2}}{4}(1 + 3\cos{\beta}) + (\Delta_{\bot,Z}^{2} + \Delta_{\parallel}^{2})\cos{\beta} - \frac{\Delta_{\parallel} \Delta_{\bot,Z}}{4} (2 + 8 \cos{\beta})) & \beta > \beta^{*}\\
		\end{cases}
\end{equation}
\end{widetext}
for a critical tip angle $\beta^{*}$ 
\begin{equation}
	\cos{\beta^{*}} = \frac{-\Delta_{\bot}^{2} + \Delta_{\bot,Z} \Delta_{\parallel}}{\Delta_{\bot}^{2} + 2 \Delta_{\bot,Z}\Delta_{\parallel}}
\end{equation}
In the limit where $\Delta_{\bot} = \Delta_{\bot,Z}$, this reduces to the expressions in Ref.\cite{Dmi.14}.
This limit is used for the fit in Fig.\ref{fig:tip_angle_shifts}.

\end{document}